\documentclass[aps,pra,tightenlines,twocolumn,showpacs,groupedaddress,superscriptaddress,longbibliography]{revtex4-1}

\usepackage[english]{babel}
\usepackage{float}
\usepackage{graphicx}
\usepackage[caption=false]{subfig}
\usepackage{amsmath}
\usepackage{amsfonts}
\usepackage{amssymb}
\usepackage{mathbrack}
\usepackage{mathtools}

\begin{document}

\title{Scaled Tight-Binding Crystal}

 \author{Peter Schmelcher}
  \email{Peter.Schmelcher@physnet.uni-hamburg.de}
 \affiliation{Zentrum f\"ur Optische Quantentechnologien, Fachbereich Physik, Universit\"at Hamburg,
 Luruper Chaussee 149, 22761 Hamburg, Germany}
 \affiliation{The Hamburg Centre for Ultrafast Imaging, Universit\"at Hamburg, Luruper Chaussee 149, 22761 Hamburg, Germany}

\date{\today}

\begin{abstract}
  The concept of local symmetry dynamics has recently been used to demonstrate the
evolution of discrete symmetries in one-dimensional chains leading to emergent periodicity. 
Here we go one step further and show that the unboundedness of this dynamics 
can lead to chains that consist of subunits of ever increasing lengths which results
in a scaled chain. Mapping this scaled chain onto a corresponding tight-binding 
Hamiltonian we investigate its spectral and transmission properties. Varying the
off-diagonal coupling the eigenvalue spectrum shows different branches with characteristic
transitions and peaks in the corresponding density of states. The fluctuations of the energy levels exhibit a hierarchy of minigaps  
each one accompanied by a characteristic sequence of energy spacings. We develop a 
local resonator model to describe the spectral properties and gain a deeper understanding
of it in the weak to intermediate coupling regime. Eigenstate maps together with 
the inverse participation ratio are used to unravel the characteristic (de-)localization 
properties of the scaled chain with varying coupling strength. Finally we probe
the energy-dependent transmission profile of the scaled chain.
\end{abstract}

\maketitle

\section{Introduction} \label{sec:introduction}

\noindent
The structure, design and applications of materials that follow a certain order principle
represents a central theme in modern quantum physics \cite{Sethna03,EdNatPhys,Gross96}.
Symmetries play in this context a pivotal role since they provide an important characteristic for
the classification and description of the systems under investigation. A prime example are periodic crystals based 
on the existence of a discrete translation symmetry that provides us with the Bloch theorem
and the celebrated concept of band structure analysis \cite{Ashcroft76,Singleton01} with vast applications
in modern material science. Quasicrystals \cite{Shechtman84,Suck02,Macia09,Macia21},
on the other hand side, fall into the gap between perfectly periodic crystals and disordered structures.
Quasiperiodic order adds new categories to the spectral classification chart, such as singular continuous
energy spectra and lattice Fourier transforms, and lead to novel physical properties emerging from the
fractal nature of their energy spectra. Opposite to periodic crystals quasicrystals are not based on
global symmetries, such as a discrete translation group, but typically exhibit a plethora of local symmetries \cite{Morfonios14}
embedded into their self-similar structure.

Structures built on basis of the concept of local symmetries, i.e. symmetries that hold only in a limited domain of space,
are very well-suited to further fill the above-mentioned gap between global order and disorder.
Indeed, several recent works \cite{Kalozoumis14a,Kalozoumis14b,Morfonios17,Schmelcher17,Zambetakis16}
have been focusing on the development of a theoretical framework of the impact of local symmetries for both
continuous and discrete systems.
Among others, it has been demonstrated that local symmetries lead to invariant non-local currents which
allow for a generalization of the Bloch theorem \cite{Kalozoumis14a}. Sum rules imposed on these invariants can
serve as a tool to classify resonances in wave scattering \cite{Kalozoumis14b}. These invariants and the corresponding
control of local symmetries have been detected in lossy acoustic waveguides \cite{Kalozoumis15} and were observed
in coupled photonic wave guide lattices \cite{Schmitt20}. Systematically introducing more and more of local
symmetries into one-dimensional disordered finite chains has been demonstrated to enhance the corresponding
transfer efficiency across the chain \cite{Morfonios20}. An important aspect of the presence of local symmetries
in the strong coupling regime is the 'formation' of so-called local resonators on which the localization
of eigenstates take place. This characteristic was developed and used in ref.\cite{Roentgen19} to analyze the eigenstate
properties and edge state appearance for quasiperiodic chains of different spectral category.

Inspired by the substitution rules used to generate quasiperiodic chains very recently the concept of
local symmetry dynamics (LSD)
has been put forward \cite{Schmelcher23} (see also \cite{Morfonios14}) to obtain one-dimensional lattices with
a plethora of local symmetries that do not belong to the periodic or quasiperiodic case. 
The idea here is to generate a lattice by applying successive reflection operations on an initial seed
of the lattice consisting of a finite number of sites. The rules generating the LSD
can be manifold, but the first case explored in ref.\cite{Schmelcher23} are the so-called $n:m$ rules.
$n$ and $m$ indicate the number of sites of the lattice involved in the reflection operations and 
are applied alternatingly in the course of the LSD. It has been shown that 
the such created one-dimensional lattice shows emergent periodic behavior, i.e. it consists of a
transient whose length depends on the concrete values of $n,m$ followed by a subsequent
periodic behavior. By construction, the local symmetries of this lattice are strongly overlapping.
A spectral analysis of the tight-binding (TB) realization of the $n:m$ LSD chains demonstrated the control possibilities
of the localization properties of the eigenstates by the nested local symmetries.

In the present work we go one step further and establish a type of rule which does not possess emergent
periodicity but leads to a scaling behavior of the chain. As a consequence we have similar repeating units along the
lattice but, in each step, they are stretched with respect to their lengths i.e. the number of sites
is correspondingly increased. We perform a detailed spectral analysis of a TB implementation of the scaled chain.
The eigenvalue spectrum shows a distinct transition from two to three branches and finally to 
a single branch with increasing off-diagonal coupling strength. The fluctuations of the energy level spacings
cover several orders of magnitude and exhibit minigaps. A density of state analysis shows a strongly peaked
behavior at the crossover points of the branches. We develop a local resonator model  which allows us to
interpret and understand this spectral behavior. Our eigenstate analysis demonstrates the unique localization
properties of this scaled chain and in particular their variation with changing coupling strength. 
Finally we investigate the energy-dependent
transmission profile by attaching leads to the scaled chain. It shows a transition from few to many isolated complete transmission
peaks and finally, for smaller values of the coupling, we observe 
a decreasing spectral transmission window with an irregular fluctuating behavior.

This work is structured as follows. In section \ref{sec:lsdchain} we introduce our LSD rule and the resulting
scaled chain and map it onto a TB Hamiltonian. Section \ref{sec:energies} presents an analysis of the energy
eigenvalue spectrum of scaled chains including the energy spacing distributions and density of states.
In section \ref{sec:model} we develop the local resonator model which offers a deeper
understanding of the spectral properties and we compare it to the TB results. Section \ref{sec:estat}
provides an analysis of the eigenstates including their localization behavior. The transmission properties
of our scaled chain are explored in section \ref{sec:transm} with varying coupling strength. Finally,
in section \ref{sec:concl} we present our conclusions.

\section{The scaled LSD chain: setup and Hamiltonian} \label{sec:lsdchain}

The local symmetry dynamics (LSD) represents a concept which allows us to generate lattices with local symmetries
starting from a given initial condition, i.e. from an initial finite segment of a lattice. One important way to 
achieve this is to perform reflection operations of a certain domain size at the end of a given finite lattice.
In ref.\cite{Schmelcher23} the special case of the $n:m$ rules, where $n,m$ stand for the number of sites
to be reflected alternatingly, has been investigated: it provides us with emergent periodicity in the sense
that a spatially evolving transient is followed by a periodic behavior of the lattice. By construction, the resulting lattice
exhibits a plethora of local overlapping symmetries.

Employing a symbolic code we focus here on the rule $n,(n+1),(n+2),(n+3),...$
which represents an LSD with monotonically increasing sizes of the
reflection domains. The initial seed has to be $n$ elements and we use here $n=2$ i.e. the seed $AB$. As an example, we
provide the tenth generation of the application of this rule which reads as follows

\begin{eqnarray} \nonumber
AB |_2 BA |_3 ABB |_4 BBAA |_5 AABBB |_6 BBBAAA \\ \nonumber
|_7 AAABBBB |_8 BBBBAAAA |_9 AAAABBBBB\\ \nonumber
|_{10} BBBBBAAAAA |_{11} AAAAABBBBBB \\ \nonumber
|_{12} BBBBBBAAAAAA |_{13} AAAAAABBBBBBB
\end{eqnarray}

where $|_k$ stands for the reflection operation exerted on $k$ sites to the left of its position.
Obviously, our rule leads to a scaling behavior in the sense that we have alternating sequences of
$A$ and $B$ sites whose lengths increase with increasing generation of the chain. Alternatively,
this can be noted as $1A,2B,2A,4B,4A,6B,6A,....2nB,2nA$ which amounts to a total length of $N = 1 + 2n(n+1)$.
We therefore call this chain a scaled chain (SC). 

In order to explore the spectral and transmission properties of the SC we map it onto a corresponding
TB Hamiltonian \cite{Goringe97}.
We hereby assume a constant off-diagonal coupling $t$ between nearest neighbors $\langle i,j \rangle$
of a discrete chain of length $N$ with sites $\{i|i=1,...,N\}$. The corresponding on-site energies
$\epsilon_i$ follow the LSD of the SC. Specifically we will use in the following the values
$\epsilon_A = 1.0, \epsilon_B = 2.0$ for the sites of type $A,B$, respectively. Our TB Hamiltonian
reads therefore as follows

\begin{equation}
{\cal{H}} = \sum_{i=1}^{N} \epsilon_i |i \rangle \langle i| + \sum_{ \langle i,j \rangle} t |i \rangle \langle j| 
\label{eq1}
\end{equation}

Note that we are using open boundary conditions for the SC allover this work.

\section{Energy eigenvalue spectra} \label{sec:energies}

\begin{figure}[!htbp]
\center
\includegraphics[width=\linewidth]{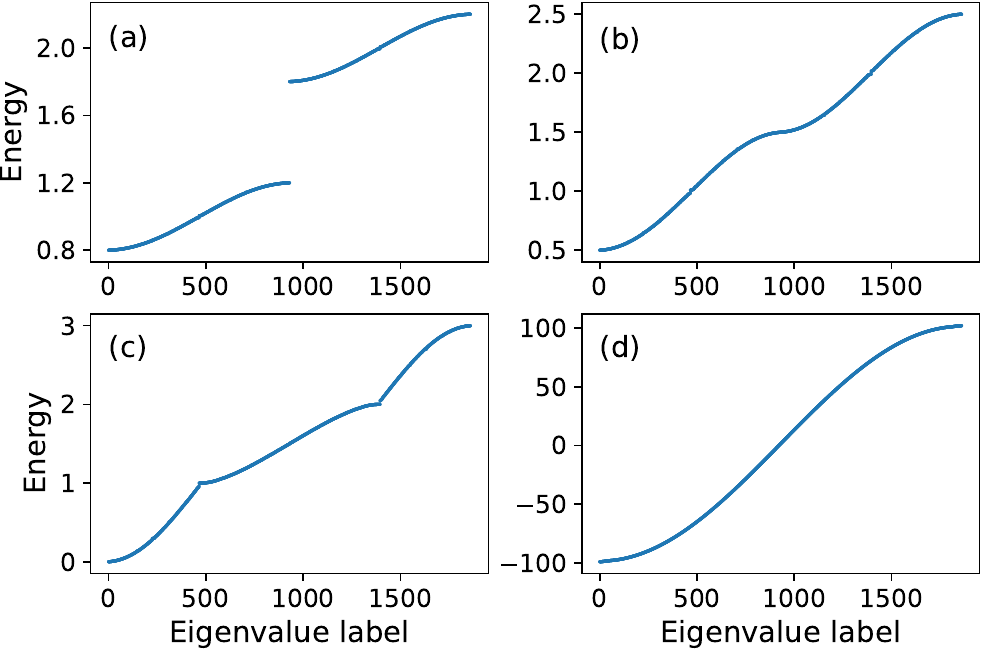}
\caption{Spectrum of the energy eigenvalues for a SC. The onsite values
used are $\epsilon_A = 1, \epsilon_B = 2$ and the off-diagonal coupling is varied according
to $0.1,0.25,0.5,50.0$ from (a) to (d), respectively. Open boundary conditions and the length
of the chain is $1861$.}
\label{Fig1}
\end{figure}

Let us analyze in this section the energy eigenvalue spectra of the SC for varying off-diagonal $t$ from
weak to strong couplings. First we inspect the global spectral behavior and subsequently
the fluctuations in terms of the energy level spacing will be discussed. We hereby focus on a
SC of length $1861$ which corresponds to subchains of purely $A$ or $B$ sites of maximal lengths $30$.
For $t=0$ we have only the two highly degenerate eigenvalues $\epsilon_A =1.0$ and
$\epsilon_B = 2.0$. Switching on the coupling we observe in Fig.\ref{Fig1}(a) for $t=0.1$ an energetically lower and upper
branch of the spectrum (we refrain from using the terminology of a band, due to the non-periodic structure
of our chain) which are separated by an energetical gap. Increasing the value of $t$ the size of this gap decreases
until at $t \approx 0.25$ the gap closes, which can be observed in Fig.\ref{Fig1}(b). Then, the low and high
energy branch are connected while possessing a similar behavior of their slopes with varying energy.
With further increasing coupling strenght $t$ a third branch appears and persists for a broad range of values
of $t$. The crossovers between the branches is characterized by a cusp. This can be observed in
Fig.\ref{Fig1}(c) for $t=0.5$ where the intermediate energy branch occupies, like the low
and high energy branch, a substantial part of the spectrum. With increasing value of the coupling the intermediate
energy branch widens until at $t \approx 30$ (not shown here)
it has taken over almost all of the spectrum. Fig.\ref{Fig1}(d) shows
the spectrum for $t=50$ where only a single branch has survived. In this case we are close to the limit of
a negligible on-site energy compared to the large coupling value, which yields a single branch ($t \rightarrow \infty$)
with the spectrum given by $E_m = 2 t \cos (\frac{m \pi}{N+1})$ with $1 \le m \le N$ \cite{Kouachi06,Kulkarni99,Willms08}.

\begin{figure}[t]
\centering
\includegraphics[width=\linewidth]{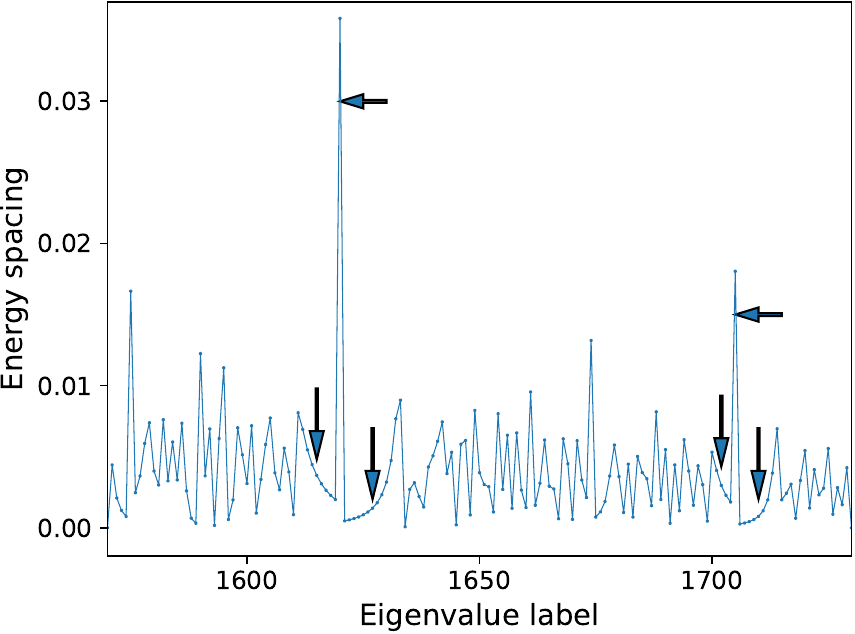}
\caption{A finite spacing sequence (levels 1570 to 1730) of the energy eigenvalue spectrum for a SC.
Arrows indicate two of the positions of the minigaps and four of the small spacing sequences located on corresponding arcs.
The onsite values are $\epsilon_A = 1, \epsilon_B = 2$ and the off-diagonal coupling is $1.0$.
Open boundary conditions and the length of the chain is $1861$.}
\label{Fig2}
\end{figure}

The above discussion relates exclusively to the envelope or mean behavior of the spectrum. Let us now address the 
fluctuations of the energy levels which are well-characterized by the spacing of the energy levels. Fig.\ref{Fig2}
shows the spacing of the energy levels in a window of the spectrum between the energy levels $1570$ to $1730$.
Notably we observe that the energy spacing covers several orders of magnitude. Interdispersed into seemingly
irregular oscillations there is two types of prominent features. First we encounter minigaps in the spectrum corresponding to
well-isolated distinguished peaks in Fig.\ref{Fig2}. Second, before and after those peaks we observe sequences of very small
energy spacings lying on arcs. Examples for both features are indicated by arrows in Fig.\ref{Fig2}. Their origin will become 
clear in the context of the local resonator model to be developed in the next section.

\begin{figure}[!htbp]
\centering
\includegraphics[width=\linewidth]{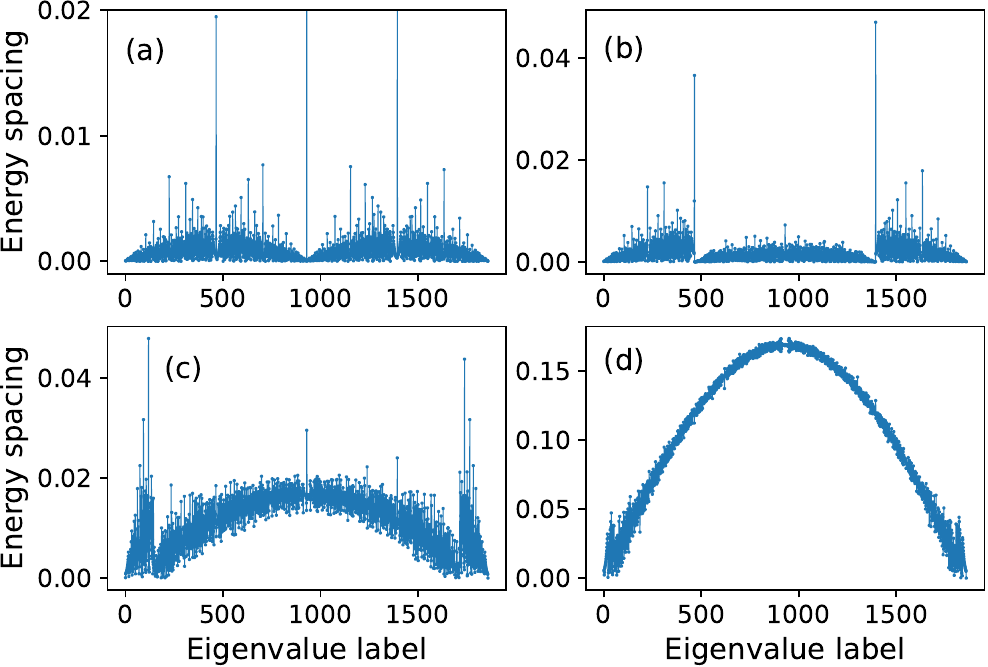}
\caption{Spacing of the energy eigenvalue spectrum for a SC. The onsite values
are $\epsilon_A = 1, \epsilon_B = 2$ and the off-diagonal coupling is varied according
to $0.2,0.5,5.0,50.0$ from (a) to (d), respectively. Open boundary conditions and the length of the chain is $1861$.}
\label{Fig3}
\end{figure}

In Fig.\ref{Fig3} the energy spacing is shown for the complete eigenvalue spectrum for four different values
of the coupling i.e. for $t = 0.2,0.5,5.0,50.0$ in subfigures (a-d), respectively. For $t=0.2$ 
the eigenvalue spectrum is still gapped which reflects itself in a dominant peak of the spacing distribution at 
the position $930$, whose value is out of the scale provided in Fig.\ref{Fig3}(a).
Left and right to this central peak there are two broad subdistributions
with strongly fluctuating values for the spacings. On top of these subdistributions there is a central dominant
peak as well as a number of additional prominent peaks which correspond to the above-mentioned minigaps
in the spectrum. Those two branches of the spacing distributions are single humped and show monotonically
decreasing spacings towards their edges.

\begin{figure}[t]
\center
\includegraphics[width=\linewidth]{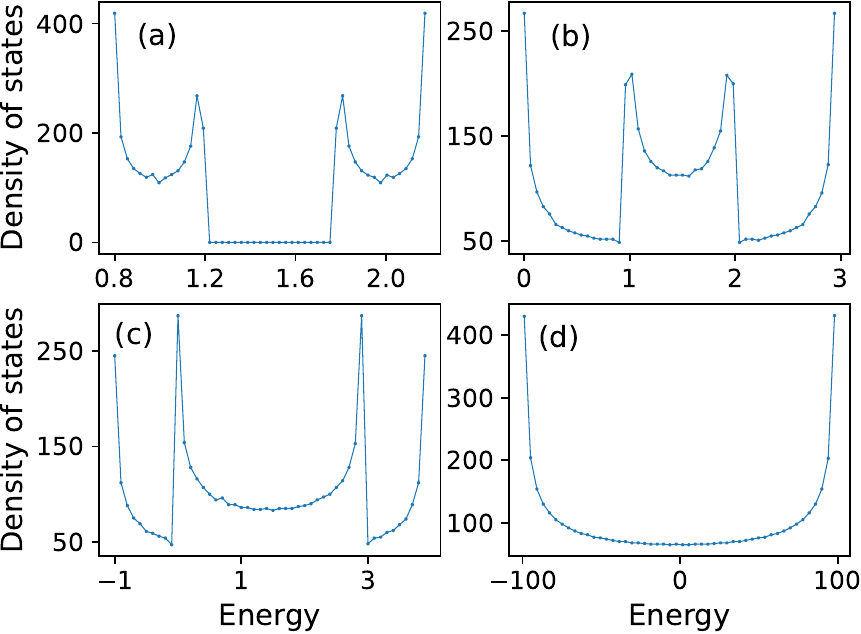}
\caption{Density of states for a SC. The onsite values
are $\epsilon_A = 1, \epsilon_B = 2$ and the off-diagonal coupling is varied according
to $0.1,0.5,1.0,50.0$ from (a) to (d), respectively. Open boundary conditions and the length of the chain is $5101$.}
\label{Fig4}
\end{figure}

Fig.\ref{Fig3}(b) corresponds to the case $t=0.5$ for which the eigenvalue spectrum possesses three branches (see Fig.
\ref{Fig1}(c)). Here the energy spacing distribution possesses also three distinct regions with abrupt transitions
between them: the position of the transition points correspond to the positions of the cusps of the energy spectrum.
At these positions of the cusps dominant peaks of the energy spacing
occur followed by a collapse of the spacing behavior with further increasing degree of excitation in the spectrum. Overall
the three branches encountered consist of two narrow semi-humps connected to the edges of the distribution and a complete
broad hump in the center of the
spacing distribution. With increasing coupling strength $t$ the central single-humped branch of the 
spacing distribution expands and finally represents the complete distributions. On this pathway the central branch
bends upwards as is clearly visible in Fig.\ref{Fig4}(c,d) for $t=1.0,50.0$ implying that the spectrum of the spacing
values possesses an increasing lower bound with increasing coupling strength. Finally, for $t=50.0$ in Fig.\ref{Fig4}(d)
the spacing distribution is already rather similar to the one expected from the off-diagonal only case: it possesses
a narrow width and only small fluctuations around the spacing values belonging to the spectrum
$E_m = 2 t \cos (\frac{m \pi}{N+1})$ with $1 \le m \le N$.

Let us conclude this section by analyzing the energetical density of states (DOS) belonging to the SC for varying coupling
strength $t$. For a periodic crystal of monomers the DOS can be obtained analytically to 
${\cal{N}}(E) \propto \frac{1}{\sqrt{1-(\frac{E}{2t})^2}}$ possessing two singularities at the band edges and in
between a smooth decrease followed by a corresponding increase. Fig.\ref{Fig4}(a) shows the case $t=0.1$ for which
a sizeable gap between a low and high energy branch of the eigenvalue spectrum exists (see discussion of
Fig.\ref{Fig1}). Correspondingly we observe four pronounced peaks for ${\cal{N}}(E)$ at the edge points of those two branches.
In between the first two and the second two peaks a smooth decrease followed by a corresponding increase is
encountered. The gap in the corresponding spectrum shows here up, of course, as a region of zero valued ${\cal{N}}(E)$. 
Fig.\ref{Fig4}(b) presents the DOS for $t=0.5$. Here the eigenvalue spectrum consists of three branches (see Fig.\ref{Fig1}(c))
and we observe two edge localized peaks and two peaks localized around the center of the DOS. The latter correspond
to the positions of the cusps in the eigenvalue spectrum. Note that abrupt transitions occuring for the left and right
side of the second and third peak of the DOS, respectively. For $t=1.0$ (Fig.\ref{Fig4}(c)) the central branch of the
DOS has widened i.e. the corresponding central peaks have been moving towards the edges of the DOS thereby maintaining
their narrow character. Finally, for $t=50.0$ only two peaks remain and are edge localized, as to be expected from
the single branch case of an (approximately) off-diagonal only TB Hamiltonian.

\section{Local resonator model}\label{sec:model}

In order to develop a profound understanding of the above-discussed features of the eigenvalue spectrum, we
develop a so-called local resonator model (LRM) for our SC. This model is inspired by the
local symmetry theory of resonator structures developed in ref.\cite{Roentgen19}. In the latter work a quantitative analysis
of the localization behavior of the eigenstates for strong and intermediate contrast has been provided for
aperiodic binary chains based on substitution rules thereby focusing on the quasiperiodic case.

Our SC $1A,2B,2A,4B,4A,6B,6A,....2nB,2nA$ consists of a sequence of purely $A$ and $B$ subchains of
increasing length with increasing size of the SC. We will call these subchains local resonators.
Starting out with small values for the off-diagonal coupling $t$ we model the SC as a superposition
of the spectra of these local $A$ and $B$ resonators. This is motivated by the fact that for zero coupling
the resonators exhibit a degenerate spectrum which is splitted for small but finite coupling whereas the
coupling between two different resonators is suppressed due to the substantial difference of their on-site energies
$\epsilon_A = 1.0$ and $\epsilon_B = 2.0$. Employing open boundary conditions we therefore have the sequence 
of spectra for the local resonators as follows

\begin{eqnarray}
\label{eq2}
E^1_A = E_A\\ \nonumber
E^2_m(A) = E_A + 2 t \cos \left( \frac{m \pi}{3} \right) \hspace*{0.5cm} m = 1,2\\ \nonumber
E^2_m(B) = E_B + 2 t \cos \left( \frac{m \pi}{3} \right) \hspace*{0.5cm} m = 1,2\\ \nonumber
E^4_m(A) = E_A + 2 t \cos \left( \frac{m \pi}{5} \right) \hspace*{0.5cm} m = 1,...,4\\ \nonumber
E^4_m(B) = E_B + 2 t \cos \left( \frac{m \pi}{5} \right) \hspace*{0.5cm} m = 1,...,4\\ \nonumber
...................................\\  \nonumber
E^n_m(A) = E_A + 2 t \cos \left( \frac{m \pi}{(n+1)} \right) \hspace*{0.5cm} m = 1,...,n\\ \nonumber
E^n_m(B) = E_B + 2 t \cos \left( \frac{m \pi}{(n+1)} \right) \hspace*{0.5cm} m = 1,...,n
\end{eqnarray}

\begin{figure}[!htbp]
\centering
\includegraphics[width=\linewidth]{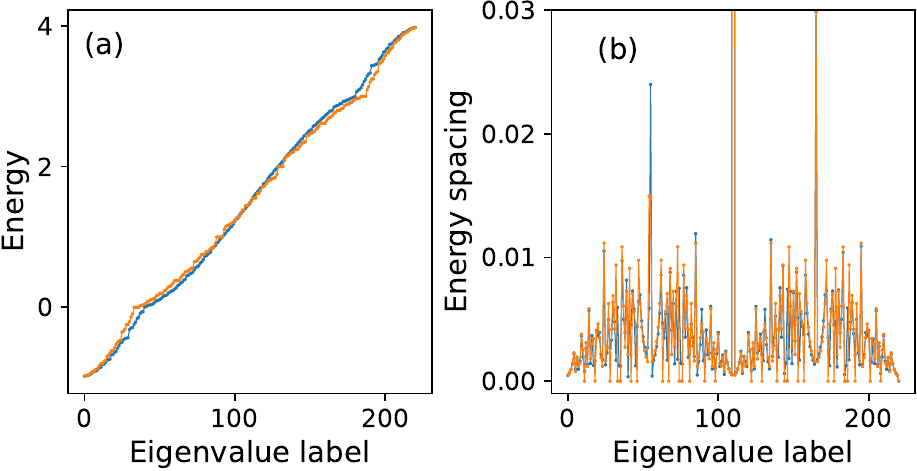}
\caption{(a) Comparison of the eigenvalue spectrum and (b) of the energy spacing of the TB chain (blue) and
its corresponding local resonator model (orange). 
The onsite values are $\epsilon_A = 1, \epsilon_B = 2$ and the off-diagonal coupling is $1.0$.
Open boundary conditions and the length of the chain is $221$.}
\label{Fig5}
\end{figure}

The corresponding eigenstates are therefore localized resonator eigenstates. Note that their superposition
is in general not an (even approximate) eigenstate of the SC, since the resonators possess different lengths.
The total spectrum of the SC within this local resonator picture is given by 

\begin{align}
\label{eq3}
\Bigl\{E_A, \Bigl\{E_A + 2 t \cos \left( \frac{m \pi}{(n+1)} \right)
|n = 2k, k \in [1,l],\\ \nonumber
 m \in [1,n] \Bigr\} \Bigr\}, \Bigl\{E_B + 2 t \cos \left( \frac{m \pi}{(n+1)} \right) |n = 2k,\\
  k \in [1,l], m \in [1,n] \Bigr\} \Bigr\} \nonumber
\end{align}

where the length of the SC, i.e. the number of sites, is given by $N = 1 + 2l(l+1)$ and the number of 
resonators in the SC is $2l+1$. The largest resonator consists of $2l$ sites. The envelope or mean behavior
of the spectrum provided by our LRM agrees with many of the envelope features of the exact SC spectrum.
E.g. it can describe the weak coupling broadening of the two bands, the subsequent closing of the energy
gap as well as the emergence and widening of the third branch with increasing coupling. Even in the $t \rightarrow \infty$
limit which corresponds to a single branch case the spectral envelope behavior can qualitatively be reproduced. 
From the LRM we can draw some general conclusions. The size of the energy gap for not too strong coupling $t$ amounts to
approximately $1-4t$ which means that the closing of the gap occurs for $t \approx 0.25$. The width of the central third
branch for $t > 0.25$ amounts to $4t-1$. Fig.\ref{Fig5}(a) shows a comparison of the LRM and the exact TB spectrum for
a SC of length $221$ and for $t=1.0$: while the overall qualitative behavior is very similar one realizes
significant deviations on smaller energy scales. 

For this reason let us now analyze the energy eigenvalue spacing
distribution of the LRM which provides us with the fluctuations of the energy levels.
Fig.\ref{Fig5}(b) shows the spacing distribution for both the LRM and the exact TB spectrum
for $t=0.1$. While the overall rough behavior is certainly similar, a closer look reveals remarkable
deviations. The fluctuations are much stronger for the LRM compared to the TB results even for these small
values of the coupling.
A substantial number of the peak spacings agree within the two approaches and in particular the 'arc-like'
accumulation of small spacings around the main peaks is reproduced well. However, an eye-catching difference
is the fact that the LRM shows zero spacings, corresponding to degeneracies in the LRM, whereas this
is not the case for the TB spectrum. Considering the spectrum of the LRM in eq.(\ref{eq3}) these degeneracies
can be shown to occur for the set of 2-tuples satisfying for a given $m$ and $k$ and for varying $\alpha$

\begin{equation}
\begin{pmatrix}
\alpha m\\
\alpha (2k+1)\\
\end{pmatrix}
\hspace*{0.5cm} \alpha \hspace*{0.2cm} \hspace*{0.2cm} \text{odd}, \alpha \in \mathbb{N}, m \le 2k, k \le l
\label{eq4}
\end{equation}

Note that the eigenstates belonging to these degenerate eigenvalues belong in general to different local resonators.
The lifting of these degeneracies in the TB spectrum is, of course, due to the interresonator coupling which is 
neglected within the LRM. An important remark is in order concerning the regime of stronger couplings $t>1$.
In section \ref{sec:energies} it has been shown that the center branch of the eigenvalue spacings
exhibits an increasing upward bending with increasing coupling $t$ (see Fig.\ref{Fig3}(c,d)), meaning that
the energy spacings systematically 'avoid small values'. This behavior cannot be described by the LRM, which
shows its inadequacy in the strong coupling regime.

\section{Localization versus delocalization of eigenstates}\label{sec:estat}

\begin{figure*}[t]
\centering
\includegraphics[width=\linewidth]{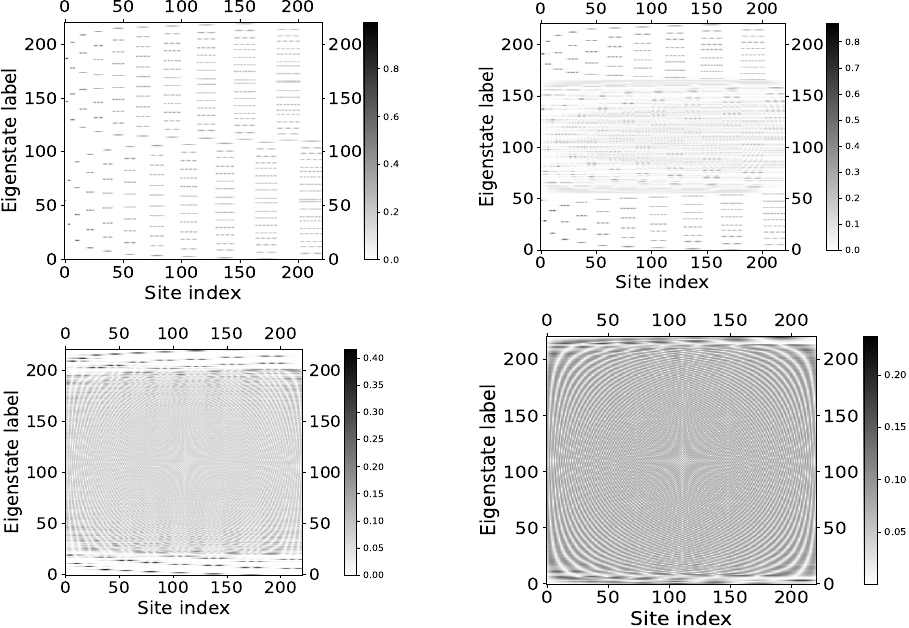}
\caption{Eigenstate maps showing the absolute values of the eigenvector components of the 
scaled TB chain. The onsite values are $\epsilon_A = 1, \epsilon_B = 2$. The off-diagonal coupling is 
$0.1$ (upper left), $0.5$ (upper right), $5.0$ (lower left) and $50.0$ (lower right).
Open boundary conditions and the length of the chain is $221$.}
\label{Fig6}
\end{figure*}

Let us now focus on the analysis of the eigenstate properties of the SC for varying coupling strength $t$ with
a particular emphasis on their localization properties. As described above in the framework of the LRM we expect for weak
coupling strengths that the local resonator picture represents a good approximation and therefore the eigenstates
should be localized within these resonators whose lengths increase monotonically along the SC.
Fig.\ref{Fig6}(a) shows the eigenstate map, i.e. a greyscale image of the magnitudes of the components
of the eigenstates with varying degree of excitation for the complete spectrum of a SC of length $221$ for $t=0.1$.

The ground state of the SC is localized on the largest $A$ local resonator of length $20$ appearing at the right end
of the SC. The first excited state of the SC is localized on the second largest $A$ local resonator to the left of the
ground state. This sequence continues (see Fig.\ref{Fig6}(a)) such that with increasing degree of excitation
the corresponding eigenstates localize on local $A$ resonators with decreasing size, thereby moving from the
right end towards the left end of the SC. This series of localized states are then intermingled with excited
states of the local resonators which adds to the localization patterns observed in Fig.\ref{Fig6}(a).
On the o.h.s. the spectrum can alternatively be described as follows. It consists of vertical series of
localized resonator eigenstates whose energy spacings decrease with increasing size of the corresponding
resonator. Series of localized resonator eigenstates of $A$ character belong to the low energy part
of the spectrum whereas those of $B$ character belong to the high energy part. According to their occurence
in the SC they are spatially shifted w.r.t. each other.

Inspecting the case $t=0.5$ in Fig.\ref{Fig6}(b) a branch of delocalized states can be observed
for intermediate energies. This branch intensifies and broadens in energy with increasing coupling
strength $t$. Indeed, for Figs.\ref{Fig6}(c,d) corresponding to $t=5.0,50.0$ the majority of eigenstates
are delocalized over the complete SC. In conclusion, we obtain a distinct transition from series of localized resonator
states for weak couplings to delocalized SC states for strong couplings which is controlled by the
interresonator coupling of $A$ (low energy) and $B$ (high energy) character.

To quantify the above observations let us determine the inverse participation ratio (IPR) for the eigenstates on
an SC with $N$ sites, which is defined as $r = \sum_{i=1}^{N} |\psi_i|^4 \in [N^{-1},1]$
for a normalized eigenvector $\sum_{i=1}^N |\psi|^2 = 1$. The maximal value for the IPR is one for an 
eigenvector localized on a single site of the chain and the minimal value $\frac{1}{N}$ is encountered for 
a state which is uniformly extended over the chain. The IPR neither depends on the position at which a 
state is localized nor is it very sensitive to the details of its distribution.

\begin{figure}[t]
\center
\includegraphics[width=\linewidth]{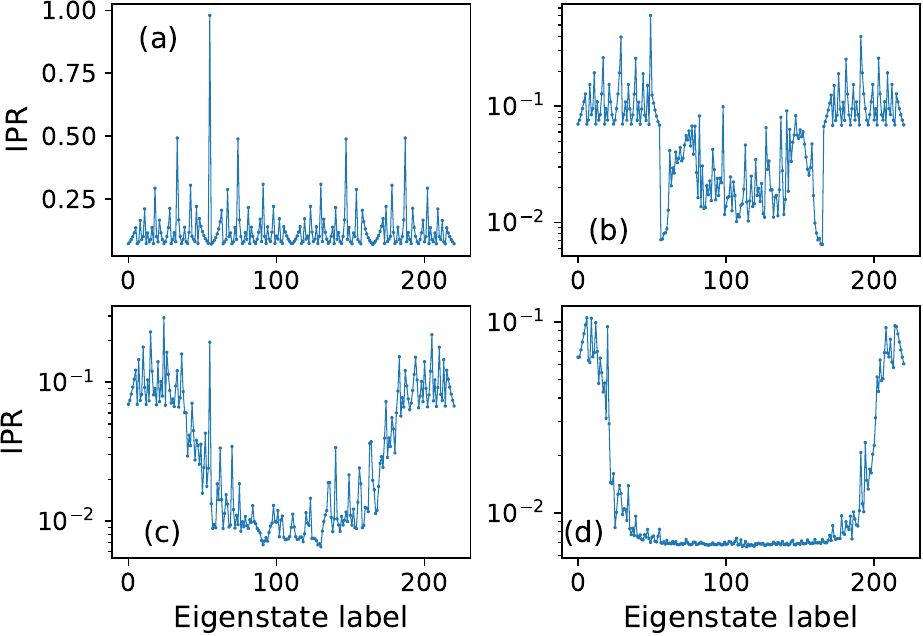}
\caption{Inverse participation ratio (IPR) for all eigenstates of the SC.
The onsite values are $\epsilon_A = 1, \epsilon_B = 2$. The off-diagonal coupling is $0.1,0.5,1.0,5.0$
for the subfigures (a)-(d), respectively. Open boundary conditions and the length of the chain is $221$.}
\label{Fig7}
\end{figure}

Fig.\ref{Fig7}(a-d) shows the IPR for all eigenstates of the SC with length $221$ for four different values of the coupling
strength $t= 0.1,0.5,1.0,5.0$. For $t=0.1$ in Fig.\ref{Fig7}(a) the IPR has a lower bound of approximately $7 \cdot 10^{-2}$
compared to the principally possible
minimum of approximately $4.5 \cdot 10^{-3}$. Most values are in the interval $[0.07,0.5]$ which indicates
that the eigenstates are very localized. The sequence of local resonator localized states with increasing energy
which have been analyzed in the context of
the discussion of Fig.\ref{Fig6} for $t=0.1$ can be seen here as a sequence of subsequent monotonically increasing
IPR values. One can also observe sequences of almost constant values of the IPR which correspond to energetically
non-neighboring eigenstates, i.e. among others involving the vertical series of localized states seen in Fig.\ref{Fig6}.

For $t=0.5$ we observe a rather sharp decrease and a regime of low-values of the IPR in the central part
of the spectrum. The latter branch of eigenstates corresponds to the delocalized states observed in the
corresponding eigenstate maps of Fig.\ref{Fig6}. With further increasing values of $t$ (see Fig.\ref{Fig7}(c,d))
the IPR value of that average central plateau decreases monotonically and the 
fluctuations on top of it systematically decrease. Finally, at $t=50.0$, apart from a small window of
states for very low and high energies, the IPR value is to a very good approximation constant
and its value is close to the minimum possible value, i.e. it is approximately $6.5 \cdot 10^{-3}$.

\section{Transmission properties of the scaled TB chain}\label{sec:transm}

In this section we explore the energy-dependent transmission through our scaled tight-binding chain. To this end
we attach two leads in the form of semi-infinite discrete chains to the left and right of our SC. Determining the
transmission follows then the standard formalism of wave function matching \cite{Zwierzycki08} which folds the infinite
leads into the SC and extends it by a single site to the left and to the right.
In the language of Greens functions this corresponds to the inclusion of the corresponding self-energy
into the Hamiltonian matrix problem: we map the closed system eigenvalue problem to a linear system
of equations with an inhomogeneity \cite{Zwierzycki08,Datta95,Ferry97}. Note that the transmission is obtained
as the absolute value squared of the $n+1$-st element of the vector obtained from the resulting linear equations.

As a result of the above procedure we have now the parameters $\epsilon_L, t_L, \epsilon_R, t_R, t_{LD},t_{DR}$
and $t, \epsilon_A, \epsilon_B$. Here, $L,R$ stand for the corresponding quantities in the left
and right lead: $\epsilon$ being the on-site energies and $t$ refering to the off-diagonal couplings,
respectively. $t_{LD}$ is the coupling of the left lead to the scattering chain and $t_{DR}$ is the
coupling of the scattering chain to the right lead, respectively. As done before we will use
$\epsilon_{A}=1.0, \epsilon_{B}=2.0$.

The leads possess due to their semi-infinite structure a continuum of $k$-values. 
We focus on the case $t_{L/R} = 1.0, \epsilon_{L/R} = 1.0, t_{LD} = 1.0, t_{DR} = 1.0$.
As a consequence the energy accessible in the left and right leads correspond to
$E = \epsilon_{L/R} + 2t_{L/R} \text{cos}(k_{L/R})$. Therefore we have $k \in [0, 2 \pi ]$
and the lead energy window is $E \in [-1.0,3.0]$. We will then vary the coupling $t$ of the SC
in the range $[0.1,50.0]$ and analyze the resulting transmission $T(E)$.
For large values of $t$ the lead energy width will be small compared to the energetical width of
the SC, whereas for small values of $t$ the opposite holds true.

\begin{figure}[t]
\center
\includegraphics[width=\linewidth]{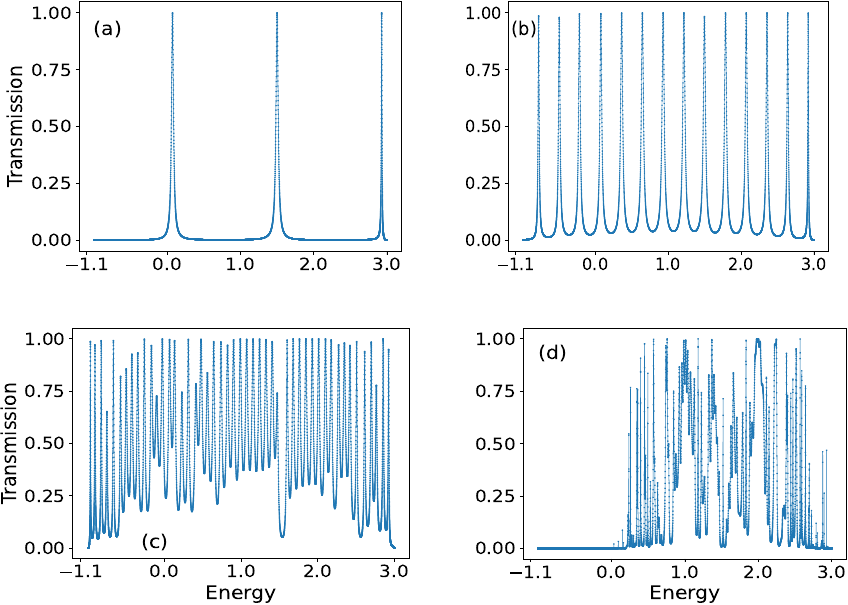}
\caption{Transmission as a function of energy for the parameters $t_{L(R)}=\epsilon_{L(R)}=t_{LD}=t_{DR}=1.0$
and for $\epsilon_A=1.0,\epsilon_B=2.0$ for the scaled TB chain. The off-diagonal coupling is $50.0,10.0,3.0,1.0$
for the subfigures (a)-(d), respectively. Open boundary conditions and the length of the chain is $221$.}
\label{Fig8}
\end{figure}

In Fig.\ref{Fig8}(a-d) we show transmission spectra for the values 
$t = 50.0, 10.0, 3.0, 1.0$. For $t=50.0$ in Fig.\ref{Fig8}(a) the energy interval of the SC is 
approximately $[-99.0,102.0]$ and, as mentioned above, the lead energy interval is $[-1.0,3.0]$.
Within this energy window the eigenstates of the isolated SC are all completely delocalized, see 
Fig.\ref{Fig6}. In this case the transmission spectrum shows three sharp peaks which are located
approximately at the energies $E = 0.0, 1.5, 3.0$. In between those peaks the transmission reaches
approximately zero, i.e. the three peaks are well-isolated. This can be explained by projecting the energy window
of the lead onto the stationary energy eigenvalue spectrum of the SC without leads only for the
above parameters: Only three eigenstates located at the energies of the transmission peaks are encountered.

If we now decrease the value of the coupling to $t=10.0$ (see Fig.\ref{Fig8}(b)), we have an energy
window of approximately $[-19.0,22.0]$ for the SC, and we observe $14$ distinct peaks.
In between this dense sequence of well-isolated peaks the transmission does not completely
decrease to the value zero due to the finite overlap of the resonances.
As a result the transmission peaks 'live' on a background of low transmission.
Again, this series of transmission peaks can be understood by projecting the energy window of the lead 
$E \in [-1.0,3.0]$ onto the energy eigenvalue spectrum of the SC for $t = 10.0$. The corresponding eigenstates of
the SC without leads are assignable to the transmission one peaks and are of completely delocalized character.
The above trend continues with e.g. for $t = 5.0$ 28 peaks (not shown) being present in the transmission
spectrum. The background transmission increases consequently. Also for $t = 3.0$ (see Fig.\ref{Fig8}(c))
we address with the lead energy window only delocalized states. The transmission spectrum shows still a series of distinct
peaks, but with a larger more irregular appearing background formed by the overlaping resonances.

For $t = 1.0$ (see Fig.\ref{Fig8}(d)) the transmission profile has changed qualitatively. Now, the lead
energy window $[-1.0,3.0]$ covers a large part of the energy eigenvalue spectrum and the corresponding
states are for low energy purely localized whereas for $E \gtrsim 0.2$ they become delocalized and
contribute to the highly irregular transmission profile. The delocalized part of the eigenstates
ends at $E \approx 3.0$ which is also the end of the transmission energy window. While the transmission
spectrum is irregular, there is, in many cases, a clear assignment to the behavior of the inverse
participation ratio: low values of the IPR correspond to delocalization which leads to high transmission values. 
For $t=0.5$ (not shown here) the transmission profile is even more fragmented into irregular series
of narrow peaks and finally, for $t=0.3$, the transmission is essentially zero in the complete spectral
window.

\section{Summary and conclusions} \label{sec:concl}

The concept of local symmetry dynamics provides a systematic pathway of generating lattices covered with
overlapping local symmetries. While this pathway has very recently been pursued to show that the class of so-called $n:m$
rules provide us with emergent periodic behavior \cite{Schmelcher23}, we show here that the local symmetry dynamics according
to the rule $n,(n+1),(n+2),(n+3),...$ leads (for $n=2$) to a scaling behavior. The resulting chain consists therefore
of the concatenation of alternating subchains of increasing lengths.

Mapping the scaled chain onto a tight binding Hamiltonian the focus of the present work has been on the analysis
of the spectral and transmission properties of this Hamiltonian. The energy eigenvalue spectrum shows with
increasing strength of the off-diagonal coupling a transition from two to three and finally to a single
branch. A closer look at the spectrum reveals minigaps accompanied by a characteristic accumulation of eigenvalues
in their neighborhood. The eigenvalue spacings exhibit a crossover from the case of a
two to a three and finally a single humped distribution. The cusps connecting the different branches go along
with sharp peaks in the corresponding density of states. Many of the features occuring for the eigenvalue spectrum
could be explained and understood via a local resonator model which applies for weak to intermediate values
of the coupling and which treats the subchains possessing the same on-site energies as independent resonators.
The spectrum and eigenstates of the complete chain are then composed of the independent resonator spectra and 
eigenstates. Indeed, an eigenstate analysis via so-called eigenstate maps demonstrates the localization of
the eigenstates in terms of resonator eigenmodes for weak couplings. The transition to delocalization with
increasing coupling strengths occurs in an expanding manner starting from the center of the spectrum and has been analyzed
via the corresponding behavior of the inverse participation ratio. Finally, we have investigated the transmission
properties of the scaled chain and found, with decreasing coupling, a transition from the case of a few regularly arranged 
isolated transmission one peaks to the case of many such peaks on an enhanced background. Finally, for 
sufficiently small coupling strengths, the transmission profile becomes highly irregular and probes the
complete delocalized eigenstate portion of the scaled chain.

The presently investigated case of a scaled chain is certainly only a specific case out of many possible
local symmetry dynamics generated chains which are asymptotically non-periodic. Indeed, one can think of
many different ways of modifying the applied rule such that a rich interplay of scaled subchains with
intermediate ones appear. The common feature of all of those chains is the presence of the extensive local
reflection symmetries of overlapping character. It is left to future investigations to possibly arrive at
a classification of those local symmetry dynamics generated lattices and the physical properties of their
resulting tight binding realizations.

Finally let us briefly address the possibility of an experimental realization of the scaled tight-binding
chain. A promising optics-based platform are evanescently coupled waveguides (see ref.\cite{Szameit10} and
in particular the references therein). Due to the extended control of the underlying optical materials
the propagation of light and the coupling among the waveguides can be varied in a wide range and the
dynamical evolution has many commons with the corresponding time evolution for a single particle
quantum system. The coupling among the waveguides can be encoded into the bulk material using
femtosecond laser pulses. A limiting factor for the current application to the scaled chain is certainly
the number of waveguides accessible which would correspondingy 
limit the largest possible resonator in the chain.

\vspace*{0.5cm}

\section{Acknowledgments} \label{sec:acknowledgments}

This work has been supported by the Cluster of Excellence “Advanced Imaging of Matter” of the Deutsche
Forschungsgemeinschaft (DFG)-EXC 2056, Project ID No. 390715994.

\end{document}